\documentstyle[aps,psfig,multicol]{revtex}

\newcommand{\ee}{\end{equation}}
\newcommand{\br}{\begin{eqnarray}}
\newcommand{\er}{\end{eqnarray}}

\newcommand{\bd}{\begin{displaymath}}
\newcommand{\ed}{\end{displaymath}}

\newcommand{\bfig}{\begin{figure}}
\newcommand{\efig}{\end{figure}}

\begin{document}
\title{Modelling of Optical Detection of
 Spin-Polarized Carrier Injection into Light-Emitting Devices}
\author{M. C. de Oliveira$^{1}$ and He Bi Sun$^{2,3}$}
\address{$^1$Departamento de F\'\i sica, CCET, Universidade Federal de S\~ao Carlos,\\
   13565-905, S\~ao Carlos, SP, Brazil.\\
$^{2}$ School of Physical Sciences, \\
$^3$Special Research Centre for Quantum Computer Technology\\
The University of Queensland, QLD 4072, Brisbane, Australia.}
\date{\today}
\maketitle
\begin{abstract}
We investigate the emission of multimodal polarized light from
Light Emitting Devices due to spin-aligned carrier injection. The
results are derived through operator Langevin equations, which
include thermal and carrier-injection fluctuations, as well as
non-radiative recombination and electronic g-factor temperature
dependence. We study the dynamics of the optoelectronic processes
and show how the temperature-dependent g-factor and magnetic field
affect the degree of polarization of the emitted light. In
addition, at high temperatures, thermal fluctuation reduces the
efficiency of the optoelectronic detection method for measuring
the degree of spin-polarization of carrier injection into
non-magnetic semicondutors.
\end{abstract}


\section{Introduction}
Advances on control of spin degree of freedom in electronic
devices has led to a strong research program in a new branch of
technology, so-called spintronics, extending the usual electronics
\cite{awschalom}. Potential applications such as spin transistors
\cite{datta} or spin memory storage devices
\cite{prinz,kikkawa,sarma} are among the main motivations for such
a technological challenge. Since spin
 decoherence time is much longer than all the relevant time scales \cite{kikkawa},
  a more ambitious proposal is to encode quantum-bits (qubits) of information, for quantum
computation protocols, on electronic spins bounded to quantum
dots\cite{divicenzo} or to silicon implanted impurities
\cite{kane}.

One main obstacle for this technological trend is to efficiently
inject (and detect) spin-polarized carriers into semiconductor
media through magnetic or semimagnetic contacts
\cite{ostreich,ostreich2,schmidt}. However, recent advances have
been reported with remarkable achievements of efficient (up to 86
\%) electrical spin-polarized carrier injection
\cite{ostreich,fiederling,ohno,jonker,jonker2} through
 a spin-aligner (spin-filter \cite{egues}) into a GaAs
 light-emitting device (LED).
 Despite the many specific details and variety of materials used as spin-aligner,
 such as BeMnZnSe \cite{fiederling}, ZnMnSe \cite{jonker},
 ferromagnetic GaMnAs epilayers \cite{ohno}, or double barrier
 resonant tunnelling diode \cite{molenkamp2,hai}
 the standard technique for detection of the efficiency of
spin-polarized carrier injection is the polarization measurement
of the device emitted light at low temperature. Selection rules
for radiative recombination process in GaAs allow a direct
relation between spin-selective injection and the emitted light
polarization. However, thermal effects such as temperature
dependence of electron g-factor \cite{oestreich3}, noise due to
thermal-light emission, as well as non-radiative carrier
recombination may blur the detected light degree of polarization,
which could cause an apparent low efficiency in spin-polarized
carrier injection at higher temperatures. Thus a detailed analysis
of thermal effects on the spin-polarized photon emission and
detection should be included in modelling the dynamic processes.

 In this paper we analyze the temperature and magnetic field dependence of the GaAs emitted
light degree of polarization,
considering a full quantum model for the generation of polarized
light in GaAs LED in the presence of a magnetic field. Effects
such as spin-polarized carrier pumping, radiative and
non-radiative recombination, as well as Zeeman splitting due to
the magnetic field  are considered in a quantum Langevin approach
\cite{sargent,fujisaki}.

There is reasonable literature on transport and noise in
conventional optoelectronic devices following the quantum Langevin
approach, such as Refs.
\cite{sargent,fujisaki,yamamoto,bjork,haug}.
Moreover, such approaches have been quite successfully applied to
description of noise in non-equilibrium quantum optical processes
\cite{milburnli,scully,gardiner} including those present in light
generation and detection. In this paper we model the quantum
processes in non-conventional spin-polarized LEDs with a
microscopic description. Particularly we extend the multimodal
light emission treatment of Ref. \cite{fujisaki} by considering
the spin degeneracy lifting when a magnetic field is applied on
the device. Such approach is quite useful for the understanding of
the relevant microscopic physical processes.

We first quantify the intrinsic degree of polarization of the GaAs
light emission, being it strongly affected by temperature effects.
The temperature dependence of the electronic g-factor is
responsible for a slight decrease of the degree of polarization,
once the decrease of the electronic g-factor with the temperature
decreases the conduction band spin-splitting sensitivity to the
magnetic field. However at higher temperatures, thermal photons
are also emitted by the GaAs device, and the intrinsic degree of
polarization decreases abruptly at a threshold temperature
($T_c$), being $T_c$ dependent on the spectral response of the
light detector as defined in Sec. V. The effect of unbalanced
spin-injection is also analyzed. We develop a quite useful
expression for the degree of polarization of the emitted light,
which shows now a dependence on the spin-aligned carrier pumping,
as well as on the radiative and the non-radiative electron-hole
recombination. Since the intrinsic polarization in GaAs is
opposite to that in spin-polarizing materials, it decreases the
net spin-injection efficiency as reported in
\cite{fiederling,jonker,jonker2}. We model the spin-polarized
carrier injection by considering the spin-aligner as a Brillouin
paramagnet\cite{furdyna}, and introduce a phenomenological
spin-polarized current density, which is dependent on the
spin-aligner layer thickness,
 the applied magnetic field and the temperature. We then describe the net polarized light
 emission due to both the intrinsic polarization of GaAs and the polarized carrier injection.

\begin{figure}
\centerline{$\;$\hskip 0
truecm\psfig{figure=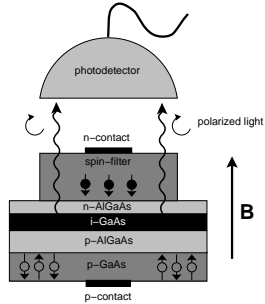,height=4cm,angle=0}} \vspace{0cm}
\centerline{ \caption{Spin-filtering device.}} \label{fig1}
\end{figure}

This paper is organized as follows. In Sec. II we begin with
describing the model for polarized-multimode photon-emission due
to radiative-recombination of spin-aligned carries in the active
layer of GaAs LEDs. In Sec. III we present the spin polarized LED
Langevin equations in a four-valence band model for the
description of polarized light generation, which includes light
and heavy hole-electron recombination. In Sec. IV we describe the
detection process. In Sec. V we analyze the influence of
temperature and magnetic filed on the generation of intrinsic
polarized light. In Sec. VI we present a quasi-equilibrium
equation for inclusion of carrier injection and non-radiative
recombination. Finally in Sec. VII we discuss enclosing the paper.

\section{Model}
The system we study is depicted in Fig. 1,  and is constituted by
a spin-aligner material layer
\cite{ostreich,fiederling,ohno,jonker} in contact with a GaAs LED,
whose emitted light is then incident on the photodetector. We
model the light emission and detection of a GaAs device only,
analyzing the intrinsic degree of polarization by setting each
sub-band in quasi-equilibrium with balanced injection of carriers.
The spin-alignment effect is phenomenologically considered by
setting unbalanced number of carriers in each spin sub-band, which
are in contact to fermionic reservoirs.
 In GaAs, the conduction band is two-fold degenerate and the
valence band is four-fold degenerate (heavy and light hole spin).
Spin degeneracy is lifted with a magnetic field, while the
light-heavy hole degeneracy is lifted by confinement
\cite{haug,ridley}. The allowed transitions are depicted in Fig.
2. Due to the selection rules, electrons with spin $-1/2$ in the
conduction band recombine with holes of spin $-3/2$ or $1/2$ in
the valence band to emit photons in right ($\sigma^+$) or left
($\sigma^-$) circular
 polarization, respectively.
Analogously  electrons with spin $1/2$ recombine with holes of spin $-1/2$ or
$3/2$ to emit photons in $\sigma^+$ or $\sigma^-$ polarization, respectively.
 In GaAs the heavy hole transition
is a factor of 3 times larger than that of the light hole.

The extended model describing polarized multimode photons and carriers in the active layer
 of the LED in the presence of a magnetic field is given by \cite{sargent,fujisaki}
\begin{eqnarray}
H&=&H_c+H_p+H_d+H_{mb}\nonumber\\
&&+H_{bath}+H_{bath-sys}+ H_M.
\end{eqnarray}
The carriers free Hamiltonian is given by
\begin{equation}
H_c=\sum_{\bf k}\left(\sum_\mu\varepsilon_{c{\bf k}\mu} c_{{\bf k}\mu}^\dagger
 c_{{\bf k}\mu}+
\sum_{\mu'}\varepsilon_{v{\bf k}\mu'} d_{-{\bf k}\mu'}^\dagger
 d_{-{\bf k}\mu'}\right),
\end{equation}
where $c_{{\bf k}\mu}$ and $d_{-{\bf k}\mu'}$ are fermionic
annihilation operators for the electron with momentum ${\bf k}$
and spin $\mu$ and the hole with momentum $-{\bf k}$ and spin
$\mu'$, respectively. The spin variables are $\mu=-1/2$, $1/2$ and
$\mu'=-3/2$, $-1/2$, $1/2$, $3/2$. $\varepsilon_{c{\bf k}\mu}$ and
 $\varepsilon_{v{\bf k}\mu'}$ are the conduction and valence band energy, respectively.
The multiphotonic process is characterized by the Hamiltonian
\begin{equation}
H_p=\sum_{l\mu\mu'}\hbar\nu_l a_{l\mu\mu'}^\dagger a_{l\mu\mu'},
\end{equation}
with $a_{l\mu\mu'}$ and $\nu_l\mu\mu'$ being the bosonic
annihilation operator  and the frequency for the photons in mode
$l$ with the polarization characterized by the allowed
spin-indexes transition $\mu$ and $\mu'$, respectively.
\vspace{-0.4cm}
\begin{figure}
\centerline{$\;$\hskip 0
truecm\psfig{figure=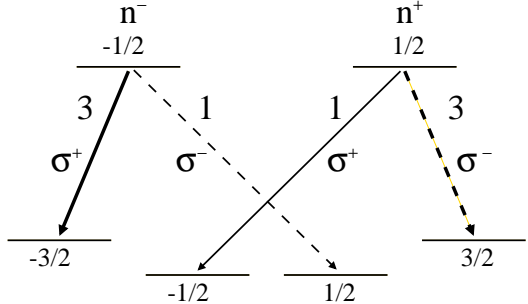,height=4cm,angle=0}} \vspace{0cm}
\centerline{ \caption{Radiative inter-band transitions allowed in
GaAs.} } \label{fig1}
\end{figure}

The dipole interaction is given by
\begin{equation}
H_d=\sum_{l{\bf k}\mu\mu'}\hbar\left(g_{l{\bf k}\mu\mu'} d_{-{\bf k}\mu'}^\dagger
 c_{{\bf k}\mu}^\dagger a_{l\mu\mu'}+H.c.\right),
\end{equation}
where $g_{l{\bf k}\mu\mu'}$ is the dipole coupling constant.
Notice that $\varepsilon_{c{\bf k}\mu}$, $\varepsilon_{v{\bf
k}\mu'}$ and $g_{l{\bf k}\mu\mu'}$ are already renormalized to
include the many-body interaction $H_{mb}$ (carrier-carrier
scattering) in a mean-field approximation \cite{sargent}. For the
direct radiative recombination in GaAs it is sufficient
 to consider
$\varepsilon_{c{\bf k}\mu}$ and $\varepsilon_{v{\bf k}\mu'}$ in a
parabolic band structure, such as $\varepsilon_{c{\bf
k}\mu}=\frac{\hbar^2k^2}{2m_e}+\varepsilon_g$ and
$\varepsilon_{v{\bf k}\mu'}=\frac{\hbar^2k^2}{2m_h}$, where $m_e$
is the conduction-band effective electron mass and
$m_h=m_{hh},m_{lh}$ is the effective mass for the heavy and light
hole, respectively; $\varepsilon_g$ describes the renormalized
band gap. To simplify the equations we have included the following
zero-rate (forbidden) transition matrix elements, $g_{l{\bf
k}-\frac 1 2-\frac 1 2}=g_{l{\bf k}-\frac 1 2\frac 3 2}=g_{l{\bf
k}\frac 1 2\frac 1 2} =g_{l{\bf k}\frac 1 2-\frac 3 2}\equiv 0$.

Let us choose a general orientation for the magnetic field and
analyze later what transitions are allowed in the Faraday
configuration, where the field is perpendicular to the layers of
the device (along z-axis) as shown in Fig. 1. The action of the
magnetic field over the device is described by the Zeeman
hamiltonian as \cite{wu}
\begin{eqnarray}\label{mag}
H_M&=&\mu_B {\bf B}\cdot \sum_{\bf k} \left(\sum_{\mu\nu}{\cal G}_e{\bf S}_{c\mu\nu}
c_{{\bf k}\mu}^\dagger
 c_{{\bf k}\nu}+\sum_{\mu'\nu'}{\cal G}_h{\bf S}_{v\mu'\nu'}
d_{-{\bf k}\mu'}^\dagger
 d_{-{\bf k}\nu'}\right),
\end{eqnarray}
where  $\mu_B$ is the Bohr magneton, ${\cal G}_{e(h)}$ is the
electron (hole) Land\'e $g$-factor and ${\bf S}_c$ and ${\bf S}_v$
are spin 1/2  matrix for electrons and spin 3/2 for holes,
respectively. Besides lifting the spin degeneracy by introducing
the Zeeman splitting, the magnetic field also induces spin-flip
between carriers sub-bands. Although magnetic fields above $1$ T
are considered in this paper, since we are only interested in a
qualitative view of the optical transitions close to the band edge
we simplify the model by not taking into account Landau levels
quantization.

In our model, the reservoir is constituted by three terms, one for
the photonic modes and the other two for electrons and holes. The
corresponding Hamiltonian terms ({${\cal H}_{bath}$} and {${\cal
H}_{bath-sys}$}) are conveniently eliminated in a Markovian
approximation for the reduced dynamics of the device
\cite{milburnli}. The photonic reservoir is assumed in a thermal
distribution, while the carriers reservoir are considered in
quasi-Fermi-Dirac distributions, where the carriers are in
equilibrium in each sub-band, but not between two of them.

\section{Spin polarized LED Langevin Equations}
%
%
 Here we consider the dynamics of the dipole
operator and for the photon number operator. The interaction with
the carrier reservoir is considered in the Langevin approach,
which includes fluctuations in the carriers and photon
populations. The Langevin equations for the dipole operator
($\sigma_{\bf k}^{\mu\mu'}=d_{-{\bf k}\mu'}c_{{\bf k}\mu}e^{i\nu_l
t}$)
 and for the photon annihilation operator ($A_{l\mu\mu'}=a_{l\mu\mu'}e^{i\nu_l t}$) describing the LED in a
 microscopic scale are given by
\begin{eqnarray}\label{dip}
\frac{d}{dt}\sigma_{\bf k}^{\mu\mu'}&=&-\frac{i}{\hbar}\left(\varepsilon_{c{\bf k} \mu}+\varepsilon_{v{\bf k} \mu'}-i \hbar \gamma
-\hbar \nu_l\right) \sigma_{\bf k}^{\mu \mu'}\nonumber\\
&&-i\sum_l g_{l{\bf k} \mu \mu'}(1-n_{e{\bf k}}^\mu-n_{h-{\bf k}}^{\mu'}) A_{l\mu\mu'} \nonumber\\
&&-\frac{i}{\hbar}\mu_B {\bf B}\cdot ({\cal G}_e\sum_\nu {\bf S}_{c\mu\nu} \sigma_{\bf k}^{\nu\mu'}+
{\cal G}_h \sum_{\nu'} {\bf S}_{v\mu'\nu'} \sigma_{\bf k}^{\mu\nu'})+F_{\sigma_{\bf k}}^{\mu\mu'}
\end{eqnarray}
and
\begin{eqnarray}\label{field}
\frac{d}{dt} A_{l\mu\mu'}&=&[-\frac{\kappa_l^0}{2}+i(\nu_l-\Omega_l)]A_{l\mu\mu'}\nonumber\\
&&-i\sum_{{\bf k}}g^*_{l{\bf k}\mu\mu'} \sigma_{\bf k}^{\mu\mu'}+F_l.
\end{eqnarray}
In these equations $\gamma$  is the dipole dephasing rate and $\kappa_l^0$ is the field decay rate, while
$F_{\sigma_{\bf k}}^{\mu\mu'}$ and $F_l$ are the fluctuation terms for the carriers and the field,
 respectively. In Eq. (\ref{field}) $\Omega_l$ is the passive-cavity (active layer) frequency \cite{sargent}.

Following Eq. (\ref{dip}) the magnetic field  induces spin-flip between each sub-band. However,
 choosing conveniently the Faraday configuration (magnetic field orientated along the device, ${\bf B}=B_z {\bf \widehat k}$) $S_z$ involves
only diagonal elements and the Eq. (\ref{dip}) is simplified to
\begin{eqnarray}\label{dipole}
\frac{d}{dt}\sigma_{\bf k}^{\mu\mu'}&=&-\frac{i}{\hbar}\left(\varepsilon_{c{\bf k} \mu}+\varepsilon_{v{\bf k} \mu'}-i \hbar \gamma
-\hbar \nu_l\right) \sigma_{\bf k}^{\mu \mu'}\nonumber\\
&&-i\sum_l g_{l{\bf k} \mu \mu'}(1-n_{e{\bf k}}^\mu-n_{h-{\bf k}}^{\mu'}) A_{l\mu\mu'} \nonumber\\
&&-\frac{i}{\hbar} \mu_B {B_z} ({\cal G}_e {S}^z_{c\mu\mu} \sigma_{\bf k}^{\mu\mu'}+
{\cal G}_h {S}^z_{v\mu'\mu'} \sigma_{\bf k}^{\mu\mu'})+F_{\sigma_{\bf k}}^{\mu\mu'},
\end{eqnarray}
  and no spin flip is present.

Now considering the regime where the dipole dephasing rate is much smaller than the field decay rate,
$\gamma\ll\kappa_l^0$ we can take the solution of Eq. (\ref{dipole}) in the slow varying regime
 for
the adiabatic approximation,
\begin{equation}\label{quasi}
\sigma_{\bf k}^{\mu\mu'}=\frac{i\sum_{l'} g_{l'{\bf k}\mu\mu'}
(n_{e{\bf k}}^\mu+n_{h-{\bf k}}^{\mu'}
-1)A_{l'\mu\mu'}+F_{\sigma_{\bf k}}^{\mu\mu'}}
{\gamma+i[\mu_BB_z({\cal G}_eS_{c\mu\mu}^z+{\cal
G}_hS_{v\mu'\mu'}^z)+\varepsilon_{c{\bf k}\mu}+ \varepsilon_{v{\bf
k}\mu'} -\hbar\nu_l)]/\hbar}, \ee and substituting it into Eq.
(\ref{field}) we obtain for the photon annihilation operator
\begin{equation}
\label{lang}
\frac{d}{dt}A_{l\mu\mu'}=[-\kappa_l^0/2+i(\nu_l-\Omega_l)]A_l+\sum_{l'}G_{ll'}^{\mu\mu'}A_{l'\mu\mu'}
+F_{\sigma l}^{\mu\mu'}+F_l, \ee where the polarized gain matrix
$G_{ll'}^{\mu\mu'}$ is defined as
\begin{equation}
G_{ll'}^{\mu\mu'}=\sum_{\bf k} G_{{\bf
k}ll'}^{\mu\mu'}\equiv\sum_{\bf k} {\cal D}_{l{\bf k}\mu\mu'}\;
g^*_{l{\bf k}\mu\mu'}\;  g_{l'{\bf k}\mu\mu'}\; (n_{e{\bf
k}}^\mu+n_{h-{\bf k}}^{\mu'}-1), \ee and we defined a new
fluctuation term
\begin{equation}
F_{\sigma l}^{\mu\mu'}\equiv-i\sum_{\bf k}g^*_{l{\bf
k}\mu\mu'}{\cal D}_{l{\bf k}\mu\mu'}F_{\sigma_{\bf k}}^{\mu\mu'},
\ee with
\begin{equation}
{\cal D}_{l{\bf k}\mu\mu'}=\frac{1}{\gamma+i[\mu_BB_z({\cal
G}_eS_{c\mu\mu}^z+{\cal G}_hS_{v\mu'\mu'}^z)+\varepsilon_{c{\bf
k}\mu}+ \varepsilon_{v{\bf k}\mu'} -\hbar\nu_l]/\hbar}. \ee

The photon number Langevin equation is obtained immediately from
Eq. (\ref{lang}) and reads
\begin{eqnarray}
\frac{d}{dt}n_{l\mu\mu'}&=&-\kappa_l^0n_{l\mu\mu'}+\sum_{l'}\left(G_{ll'}^{\mu\mu'}A_{l\mu\mu'}^\dagger
 A_{l'\mu\mu'}+H.c\right)\nonumber\\
&&+\left[(\sum_{\mu\mu'}F_{\sigma
l}^{\mu\mu'}+F_l)A_{l\mu\mu'}^\dagger+H.c.\right].\label{llang}
\end{eqnarray}
Eq. (\ref{llang}) explicitly shows the polarizations $\mu$ and
$\mu'$ dependence, while the  dissipative term is independent of
polarization once $\kappa_l^0=\nu_l/Q$, where $Q$ is
 the cavity (active layer) quality factor.

Correlations between distinct modes can be important, as for
example in the generation of sub-Poissonian light
\cite{yamamoto,bjork}, however for our interest here, we consider
the simple situation when correlations between distinct modes can
be neglected, and thus
\begin{eqnarray}
\langle
A_{l\mu\mu'}^\dagger(t)A_{l'\rho\rho'}(t)\rangle&=&\langle n_l\rangle\delta_{ll'}\delta_{\mu\rho}\delta_{\mu'\rho'},\\
\langle
\sigma_{{\bf k}}^{\dagger \mu\mu'}(t)\sigma_{{\bf k}'}^{\rho\rho'}(t)\rangle&=&\langle n_{e{\bf k}}^{\mu} n_{h-{\bf k}}^{\mu'}
\rangle\delta_{ll'}\delta_{\mu\rho}\delta_{\mu'\rho'},\\
\langle
\sigma_{{\bf k}}^{\mu\mu'}(t)\sigma_{{\bf k}'}^{\dagger \rho\rho'}(t)\rangle&=&\langle (1-n_{e{\bf k}}^{\mu})(1- n_{h-{\bf k}}^{\mu'})
\rangle\delta_{ll'}\delta_{\mu\rho}\delta_{\mu'\rho'},\\
\langle
n_{e{\bf k}}^\mu(t)n_{e{\bf k}'}^{\rho}(t)\rangle&=&\langle n_{e{\bf k}}^{\mu}\rangle\delta_{{\bf kk}'}\delta_{\mu\rho},\\
\langle F_{\sigma_{\bf k}}^{\dagger\mu\mu'}(t)F_{\sigma_{\bf
k}'}^{\rho\rho'}(t)\rangle&=& 2 D_{\sigma_{{\bf
k}}^\dagger\sigma_{{\bf
k}'}}^{\mu\mu'}\delta(t-t')\delta_{\mu\rho}
\delta_{\mu'\rho'}\delta_{{\bf kk}'}. \er

To determine the fluctuation terms we have to recall the
generalized Einstein relation \cite{sargent}. If the generalized
Langevin equation is given by
\begin{equation}
\frac{d}{dt}A_\mu=D_\mu+F_\mu
\ee
then the generalized Einstein relation will be
\begin{equation}
2D_{\mu\nu}=\frac{d}{dt}\langle A_\mu A_\nu\rangle-\langle D_\mu A_\nu\rangle-\langle A_\mu D_\nu\rangle
\ee
and
\begin{equation}
\langle F_\mu(t)F_\nu(t')\rangle = 2D_{\mu\nu}\delta(t-t'),
\ee
which is a manifestation of the fluctuation-dissipation theorem \cite{gardiner}.

Referring back to the Eq. (\ref{dipole}) we find the following diffusion term
\br
2 D_{\sigma_{{\bf k}}^\dagger\sigma_{{\bf k}'}}^{\mu\mu'}&=&\frac{d}{dt}\langle
\sigma_{{\bf k}}^{\dagger \mu\mu'}\sigma_{{\bf k}'}^{\rho\rho'}\rangle+2\gamma\langle
\sigma_{{\bf k}}^{\dagger \mu\mu'}\sigma_{{\bf k}'}^{\rho\rho'}\rangle\nonumber\\
&=&\frac{d}{dt}\langle n_{e{\bf k}}^\mu n_{h-{\bf
k}}^{\mu'}\rangle+2\gamma\langle n_{e{\bf k}}^\mu n_{h-{\bf
k}}^{\mu'}\rangle. \er Assuming the quasi-equilibrium condition
\begin{equation}
\frac{d}{dt}\langle n_{e{\bf k}}^\mu n_{h-{\bf
k}}^{\mu'}\rangle\ll 2\gamma\langle n_{e{\bf k}}^\mu n_{h-{\bf
k}}^{\mu'}\rangle, \ee we obtain
\begin{equation}
\langle F_{\sigma_{\bf k}}^{\dagger\mu\mu'}(t)F_{\sigma_{\bf k}'}^{\rho\rho'}(t)\rangle=
2 \gamma \langle
n_{e{\bf k}}^\mu n_{h-{\bf k}}^{\mu'}\rangle
\delta(t-t')\delta_{\mu\rho}
\delta_{\mu'\rho'}\delta_{{\bf kk}'}.
\ee

Analogously
\br
\langle F_{\sigma_{\bf k}}^{\mu\mu'}(t)F_{\sigma_{\bf k}'}^{\dagger\rho\rho'}(t)\rangle&=&
2 \gamma \langle
(1-n_{e{\bf k}}^\mu)(1-n_{h-{\bf k}}^{\mu'})\rangle\nonumber\\
&&\times
\delta(t-t') \delta_{\mu\rho} \delta_{\mu'\rho'} \delta_{{\bf kk}'}.
\er

For the light-field Langevin-force, considering the
non-correlation between modes, as is well known \cite{sargent}
\begin{equation}
\langle F_{l}^{\dagger}(t)F_{l'}(t')\rangle= \kappa_l^0
\bar{n}_0(\nu_l)\delta(t-t')\delta_{ll'}. \ee where
$\bar{n}_0(\nu_l)$ is the number of thermal photons. For the
carriers Langevin force, we obtain the time correlation \br
\langle F_{\sigma l}^{\dagger \mu\mu'}(t)F_{\sigma
l}^{\rho\rho'}(t')\rangle&=& \sum_{{\bf k} k'}g_{l{\bf
k}\mu\mu'}^* g_{l{\bf k}'\mu\mu'}{\cal D}_{l{\bf k}\mu\mu'}^*
{\cal D}_{l{\bf k}'\mu\mu'}
\langle F_{\sigma_{\bf k}}^{\dagger \mu\mu'}(t)F_{\sigma_{\bf k}'}^{\mu\mu'}(t')\rangle\nonumber\\
&=&
\sum_{{\bf k} k'}g_{l{\bf k}\mu\mu'}^* g_{l{\bf k}'\mu\mu'}{\cal D}_{l{\bf k}\mu\mu'}^* {\cal D}_{l{\bf k}'\mu\mu'}
2 \gamma \langle n_{e{\bf k}}^\mu(t)n_{h-{\bf k}}^{\mu'}\rangle\nonumber\\
&&\times
\delta(t-t') \delta_{\mu\rho} \delta_{\mu'\rho'} \delta_{{\bf kk}'}\nonumber\\
&=&\sum_{{\bf k}}|g_{l{\bf k}\mu\mu'}|^2|{\cal D}_{l{\bf k}\mu\mu'}|^2
2 \gamma \langle n_{e{\bf k}}^\mu(t)n_{h-{\bf k}}^{\mu'}\rangle\nonumber\\
&&\times \delta(t-t') \delta_{\mu\rho} \delta_{\mu'\rho'}. \er
Rewriting it in terms of the Lorentzian line-shape, $ {\cal
L}^{\mu\mu'}_{l{\bf k}}\equiv\gamma^2|{\cal D}_{l{\bf
k}\mu\mu'}|^2$, and the spontaneous emission rate into the mode
$l$ due to the transition $\mu\mu'$, $R_{sp,l}^{\mu\mu'}$, given
by
\begin{equation}\label{spont}
R_{sp,l}^{\mu\mu'}\equiv\frac{2}{\gamma}\sum_{\bf k}|g_{l{\bf
k}\mu\mu'}|^2 {\cal L}^{\mu\mu'}_{l{\bf k}} n_{e{\bf k}}^\mu
n_{h-{\bf k}}^{\mu'},
\end{equation}
we get
\begin{equation}
\langle F_{\sigma l}^{\dagger \mu\mu'}(t)F_{\sigma l}^{\rho\rho'}(t')\rangle=\langle  R_{sp,l}^{\mu\mu'} \rangle
\delta(t-t')\delta_{\mu\rho}\delta_{\mu'\rho'}.
\end{equation}
Similarly
\begin{equation}
\langle F_{\sigma l}^{\mu\mu'}(t)F_{\sigma l}^{\dagger \rho\rho'}(t')\rangle=\langle  R_{abs,l}^{\mu\mu'} \rangle
\delta(t-t')\delta_{\mu\rho}\delta_{\mu'\rho'},
\end{equation}
where the absorption rate is defined as
\begin{equation}
R_{abs,l}^{\mu\mu'}\equiv\frac{2}{\gamma}\sum_{\bf k}|g_{l{\bf k}\mu\mu'}|^2 {\cal L}^{\mu\mu'}_{l{\bf k}} (1-n_{e{\bf k}}^\mu)
(1-n_{h-{\bf k}}^{\mu'}).
\end{equation}

Neglecting $l\neq l'$ (intermode) correlations the photon-number
Langevin equation then writes as
\begin{eqnarray}
\frac{d}{dt}n_{l\mu\mu'}&=&-\kappa_l^0n_{l\mu\mu'}+\left(G_{ll}^{\mu\mu'}+G_{ll}^{*\mu\mu'}\right)n_{l\mu\mu'}\nonumber\\
&&+\left[(F_{\sigma l}^{\mu\mu'}+F_l)A_{l\mu\mu'}^\dagger+H.c.\right],
\end{eqnarray}
and noticing that
\begin{equation}
G_{ll}^{\mu\mu'}+G_{ll}^{*\mu\mu'}=R_{sp,l}^{\mu\mu'}-R_{abs,l}^{\mu\mu'},
\end{equation}
then
\begin{eqnarray}\label{numer}
\frac{d}{dt}n_{l\mu\mu'}&=&-\kappa_l^0 n_{l\mu\mu'}-\left(R_{abs,l}^{\mu\mu'}
-R_{sp,l}^{\mu\mu'}\right)n_{l\mu\mu'}\nonumber\\
&&+\left[(F_{\sigma l}^{\mu\mu'}+F_l)A_{l\mu\mu'}^\dagger+H.c.\right].
\end{eqnarray}
The steady state solution of Eq. (\ref{numer}) is readily
obtained, to give the steady average photon number in the mode $l$
\begin{equation}\label{num}
\bar{n}_{l\mu\mu'}=\frac{\langle(F_{\sigma l}^{\mu\mu'}A_{l\mu\mu'}^\dagger+H.c.)\rangle+\langle(F_{l}A_{l\mu\mu'}^\dagger+H.c.)\rangle}
{\kappa_l^0+(\langle R_{abs,l}^{\mu\mu'}\rangle-\langle R_{sp,l}^{\mu\mu'}\rangle)}.
\end{equation}
To calculate the correlations $\langle F_{\sigma l}^{\mu\mu'}(t)A_{l\mu\mu'}^\dagger(t)\rangle$
and $\langle F_{l}(t)A_{l\mu\mu'}^\dagger(t)\rangle$ we assume that
\begin{equation}\label{aex}
A_{l\mu\mu'}(t)=A_{l\mu\mu'}(t-\Delta t)+\int_{t-\Delta t}^t dt'\; \dot{A}_{l\mu\mu'}(t')
\end{equation}
where $\Delta t$ is an interval much shorter than $1/\kappa_l^0$
but much longer than the correlation time of the field reservoir
\cite{sargent}. Substituting Eq.(\ref{lang}) into (\ref{aex}) we
can calculate the above correlations, which then are given by \br
\langle F_{l}(t)A_{l\mu\mu'}^\dagger(t)+H.c.\rangle&=&\kappa_l^0 \bar{n}_0(\nu_l)\\
\langle F_{\sigma
l}^{\mu\mu'}(t)A_{l\mu\mu'}^\dagger(t)+H.c.\rangle&=& \langle
R_{sp,l}^{\mu\mu'}\rangle \er Substituting these correlations into
Eq. (\ref{num}) we finally obtain
\begin{equation}\label{num2}
\bar{n}_{l\mu\mu'}=\frac{\kappa_l^0\bar{n}_0(\nu_l)+\langle
R_{sp,l}^{\mu\mu'}\rangle} {\kappa_l^0+(\langle
R_{abs,l}^{\mu\mu'}\rangle-\langle R_{sp,l}^{\mu\mu'}\rangle)},
\end{equation}
which shows exactly how the absorption and emission rate
contribute to the steady average photon number in mode $l$.
As it is expected, $\bar{n}_0(\nu_l)$ coming from a thermal
reservoir (thermal photons) does not contribute to a specific
polarization. In the device working regime, $\langle
R_{sp,l}^{\mu\mu'}\rangle)\gg\kappa_l^0\bar{n}_0(\nu_l)$, the
radiative recombination process determines the light polarization.
However the increase of temperature may blur the light
polarization. We further analyze this point in the next section
for the measurement of the polarized light.

\section{Measurement of spin polarization by detection of emitted light}
At this point it is interesting to analyze the degree of
polarization of the emitted light as a function of the carriers
recombination. For that we will focus on the $l$-mode photon flux
$N_l$ at the photodetector (see Fig. 1), which we assume as placed
at the wall of the semiconductor active layer ``microcavity"
\cite{fujisaki}. The input-output theory
\cite{fujisaki,milburnli,collett} determines that the relation
between the
 output, input and the cavity field is given by
\begin{equation}\label{flux}
V_l^{\mu\mu'}=\kappa_l^0 n_{l\mu\mu'}-F_{\kappa,l},
\end{equation}
where $V_l^{\mu\mu'}$ is the photon flux of mode $l$ from the
cavity (active layer of the LED) and $F_{\kappa,l}$ is
 the input field fluctuation, which in our case is a thermal white noise.
 Now the relation between the emitted flux $V_l^{\mu\mu'}$ and the
detected flux $\bar{N}_l^{\mu\mu'}$ is given by
\begin{equation}\label{detec}
\bar{N}_l^{\mu\mu'}=\xi_l\langle V_l^{\mu\mu'}\rangle,
\end{equation}
where $\xi_l\equiv\xi(\nu_l)$ is the transmission coefficient of mode $l$. $\xi_l$ is related to the spectral
 response of the photodetector. A non-homogeneous-detection process reflects a structured response due to a
narrow-band photodetector. In the case of homogeneous detection,
or a broad-band detector, $\xi_l=\beta_0$ is a flat distribution
over the frequencies \cite{dereniak}. We shall consider only this
last situation. Thus the total detected photon-number is $\bar
N^{\mu\mu'}=\sum_l\bar N_l^{\mu\mu'}=\beta_0\sum\langle
V_l^{\mu\mu'}\rangle$. Since we did not consider correlations
between modes, the total detected photon-number is a summation of
the photon-number of each mode. Therefore, from now on it is
enough to consider the calculations for only one mode being the
extension for the multimodes. The electroluminescence intensity in
right ($\sigma^+$) and left ($\sigma^-$) circular
 polarization are given by $\bar N^+_l=\bar N^{-\frac 1 2 -\frac 3 2}_l+ \bar N^{\frac 1 2 -\frac 1 2}_l$
 and $\bar N^-_l= \bar N^{-\frac 1 2 \frac 1 2}_l+\bar N^{\frac 1 2 \frac 3 2}_l$, respectively.
We simplify our treatment if we consider the low injection limit
$\kappa_l^0\gg \langle
R_{abs,l}^{\mu\mu'}-R_{sp,l}^{\mu\mu'}\rangle$, where we can
rewrite Eq. (\ref{num2}) simply as
\begin{equation}
\bar{n}_{l\mu\mu'}=\bar{n}_0(\nu_l)+\langle R_{sp,l}^{\mu\mu'}/\kappa_l^0\rangle,
\end{equation}
and so, the photon flux at the detector is
\begin{equation}\label{detecflux}
\bar{N}^{\mu\mu'}_l=\beta_0 \left(\kappa_l^0\bar{n}_0(\nu_l)+R_{sp,l}^{\mu\mu'}\right).
\end{equation}

Following \cite{jonker} the spectral degree of polarization of the
detected light in mode $l$ is given by
\begin{equation}\label{pol}
{\cal P}(\nu_l)= \frac{\bar I^+ -\bar I^-}{\bar I^+ +\bar I^-}
\end{equation} where $I^\pm \equiv N^\pm_l /\xi_l$, is the light
intensity at the detector. Substituting Eqs. (\ref{flux}) and
(\ref{detec}) into (\ref{pol}) for a broad band detector, we
obtain the spectral degree of polarization in terms of the average
photon-number in mode $l$,
\begin{equation}\label{pol1}
{\cal P}(\nu_l)=\frac{\bar n_{l-\frac 1 2 -\frac 3 2}+\bar
n_{l\frac 1 2 -\frac 1 2}-\bar n_{l-\frac 1 2 \frac 1 2}-\bar
n_{l\frac 1 2 \frac 3 2}} {\bar n_{l-\frac 1 2 -\frac 3 2}+\bar
n_{l\frac 1 2 -\frac 1 2}+\bar n_{l-\frac 1 2 \frac 1 2}+\bar
n_{l\frac 1 2 \frac 3 2}} \end{equation} which is independent of
the transmission efficiency $\beta_0$. In the low injection limit
$\kappa_l^0\gg \langle
R_{abs,l}^{\mu\mu'}-R_{sp,l}^{\mu\mu'}\rangle$, Eq. (\ref{pol1})
writes
\begin{equation}\label{poldeg}
{\cal P}(\nu_l)=\frac{\langle R_{sp,l}^{-\frac 1 2 -\frac 3
2}+R_{sp,l}^{\frac 1 2 -\frac 1 2}-R_{sp,l}^{-\frac 1 2 \frac 1
2}-R_{sp,l}^{\frac 1 2 \frac 3 2}\rangle} {\langle
R_{sp,l}^{-\frac 1 2 -\frac 3 2}+R_{sp,l}^{\frac 1 2 -\frac 1
2}+R_{sp,l}^{-\frac 1 2 \frac 1 2}+R_{sp,l}^{\frac 1 2 \frac 3
2}\rangle+4\kappa_l^0\bar n_0 (\nu_l)}. \end{equation} The role of
the material dipole matrix for the degree of polarization is made
clear trough the
 spontaneous emission rate,
 $R_{sp,l}^{\mu\mu'}$ from Eq. (\ref{spont}), as well as the polarization
 dependence on the thermal photon number.
Notice that the broader is the detector spectral response the
stronger will be the counter effect of thermal photons over the
intrinsic degree of polarization. For a sufficiently broad
spectral response, as the temperature is raised the unpolarized
thermal photons become more and more important in the process,
decreasing the degree of polarization of the emitted light.

\section{Intrinsic polarization}

 Let us focus our discussion on the analysis of the
intrinsic polarization of the GaAs electroluminescence spectra as
a function of the temperature and the applied magnetic field. In
Fig. 3 we plot the electroluminescence spectra with right and left
circular polarization, for several magnetic fields ($0$, $1$, $4$
and $8$ T) with the temperature
 set to $T=4.2$ K.
To estimate it quantitatively we have assumed that the dipole
matrix elements are given by the $\bf k \cdot p$ theory in the
parabolic band model, being
\begin{equation}
g_{{\bf k}l\mu\mu'}= g_{l\mu\mu'}(0)\frac{\varepsilon_g}{\varepsilon_g+\frac{\hbar^2k^2}{2}\left(\frac{1}{m_\mu}+\frac 1{m_{\mu'}}\right)}
\end{equation}
where
\begin{equation}
g_{l\mu\mu'}(0)=\frac{i\;e p_{\mu\mu'}}{m_0\varepsilon_g},
\end{equation}
is the dipole momentum at the center of the band, with $e$ for the
electron charge, and $p_{\mu\mu'}$ for the electron momentum given
by the selection rules. All parameters are set to match optical
transitions in GaAs.
 \vspace{-0.4cm}
\begin{figure}
\centerline{$\;$\hskip 0
truecm\psfig{figure=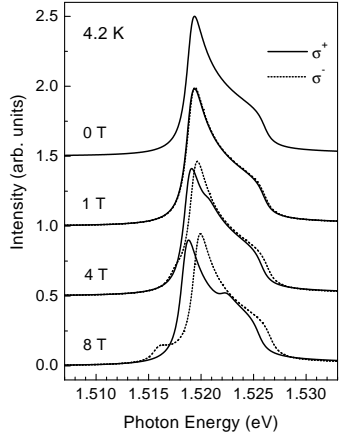,height=7cm,angle=0}} \vspace{0cm}
\centerline{ \caption{Intrinsic electroluminescence spectra of
GaAs as function of the magnetic field.} } \label{fig3}
\end{figure}

From Fig. 3 we observe that this simple parabolic band model is
reasonably good enough to give a qualitative picture of the
spectra of the
 polarized light emission, including light-hole and heavy-hole features \cite{jonker2}.
   In Fig. 3 the solid line stands for
 right-circular polarization emission, while the dotted line stands
 for left-circular polarization emission. At $B=0$ T there is no
 light polarization and both components have the same line-shape.
 As the magnetic field is increased a slight splitting of both
 spectra are noticeable and at 8 T they can be completely
 distinguished. We have observed from our calculations that the
 strongest contribution for the deformation of the polarized-light
 spectra is due to the heavy-hole feature, as it is expected \cite{fiederling,jonker}.
 Notice that some of the spectral features have opposite
 polarization, reducing thus the net light emission polarization as confirmed
 experimentally by B. T. Jonker {\it et al.}\cite{jonker2}. Those
 line-shapes can be strongly modified by the variation of the width of the GaAs
 quantum well in the AlGaAs/GaAs/AlGaAs LED, which mainly affects the energy splitting of
 the heavy and light-hole bands.
\begin{figure}
\centerline{$\;$\hskip 0 truecm\psfig{figure=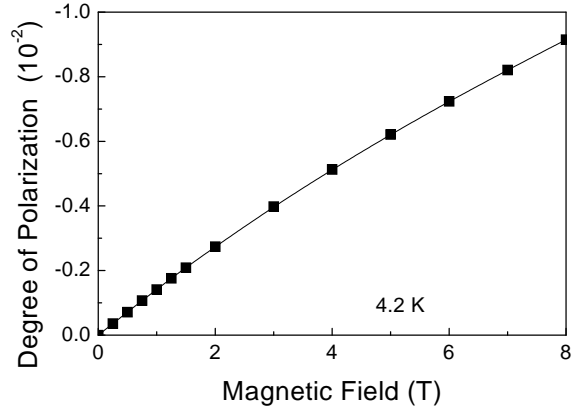,height=12cm}}
 \vspace{-5cm}
\centerline{ \caption{Intrinsic polarization degree of GaAs in
function of the magnetic field.} } \label{fig4}
\end{figure}

The intrinsic degree of polarization of the GaAs is given by
integrating (\ref{poldeg}) over the frequency range, $P=\int
d\nu_l {\cal P}(\nu_l)$. In Fig. 4 we plot the GaAs intrinsic
degree of polarization varying the magnetic field with the
temperature set to $4.2$ K. Figure 4 shows an almost linear
behavior of the degree of polarization for a weak magnetic field,
$B\le 1$ T. However as the magnetic field is increased the
polarization attains a polynomial shape. The calculated intrinsic
polarization for carrier radiative recombination corroborates
qualitatively with the experimentally measured photoluminescence
intrinsic degree of polarization for GaAs given in Ref.
\cite{fiederling} and quantitatively for electroluminescence
measurements given in Ref. \cite{ploog}.

 The variation of the intrinsic
polarization with the temperature is plotted in Fig. 5 for
 a magnetic field set to $8$ T.
The temperature dependence of the electronic g-factor is the main
 responsible by the slightly decrease of the degree of polarization shown in the figure,
once the GaAs electronic g-factor decreases with the temperature
as ${\cal G}_e=-0.44+5\times10^{-4}T$ \cite{oestreich3}, turning
the conduction band spin-splitting less sensitive to the magnetic
field. Within our model a threshold for the decrease of the
polarization is observed around $T_c=235$ K, where $T_c$ is a
critical temperature dependent on the spectral response range of
the light detector. For the present calculation we have fixed the
detector frequency range to $1$ eV, which is a reasonably good
range for detection of the central carrier radiative recombination
features. The threshold is due to thermal photons emission. At
higher temperatures thermal photons are largely emitted, washing
out the polarized emission around $1.519$ eV and the intrinsic
degree of polarization decreases abruptly as in the inset of Fig.
5. The slight increase of the polarization before the threshold at
$T_c$ is due to the fact that thermal photons start to contribute
at lower frequencies from the left side of the emission spectra
(Fig. 3) washing out first the central peak feature polarization
and then only a right-lateral feature contribution enters into the
computation of the degree of polarization.  The dependence of the
critical temperature with the detector spectral response is an
interesting issue, and is going to be addressed elsewhere. Anyhow,
besides the well known mechanisms preventing efficient spin
injection at at room temperature (see, e.g. Ref.\cite{ploog}), the
observation of spin polarized carrier injection by means of
optical polarization is also highly inefficient at those
temperatures, since thermal photons emission reduces the net
optical polarization \cite{footnote}.
 Remark that
even the $2 \%$ efficiency of spin polarized carrier injection at
room temperature observed by optical means in Ref. \cite{ploog}
was calculated by considering only lateral features of the
emission spectrum. Indeed, the net polarization calculated by
considering their whole spectrum is drastically reduced to
approximately zero, in complete agreement with our calculations
(inset of Fig. 5).

\vspace{-1cm}
\begin{figure}
\centerline{$\;$\hskip 0
truecm\psfig{figure=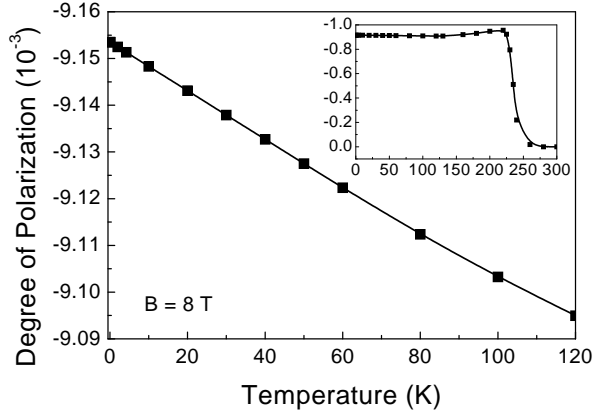,height=12cm,angle=0}}\vspace{-5cm}
 \centerline{
\caption{Decreasing of intrinsic polarization of GaAs as a
function of the temperature.} } \label{fig5}
\end{figure}

\section{Carrier pumping and non-radiative recombination}
\subsection{Carrier Langevin equation and light emission polarization rate}
 It is interesting to analyze the problem of polarized
electroluminescence if an unbalanced carrier injection is taken
into account. In such a non-equilibrium case, fluctuation effects
of carrier pumping and recombination are very important. For that
we also write a Langevin equation for the electron number operator
including carrier pumping, non-radiative recombination and
dissipative effects as well. In light emitting devices, contrarily
to laser diodes, there is very little optical feedback (if any),
and so stimulated emission and absorption can be neglected
\cite{bjork}.
Non-radiative recombination is introduced phenomenologically,
following Refs. \cite{sargent,fujisaki}. The Langevin equation for
the carrier occupation probability can be written as
\begin{equation}\label{carrier}
\frac{d}{dt}n_{e{\bf k}}^\mu=\Lambda_{e{\bf k}}^\mu(1-n_{e{\bf
k}}^\mu)-\gamma_{nr}^\mu n_{e{\bf k}}^\mu +\sum_{l\mu'}(i
g^*_{l{\bf k}\mu\mu'} A_{l\mu\mu'}^\dagger\sigma_{\bf
k}^{\mu\mu'}+H.c.)+F_{e{\bf k}}^\mu
\end{equation}
where $\Lambda_{e{\bf k}}^\mu$ is the pumping rate due to a current injection,
  $(1-n_{e{\bf k}}^\mu)$ is the pump blocking,
$\gamma_{nr}^\mu$ is the non-radiative recombination parameter
included phenomenologically, and $F_{e{\bf k}}^\mu$ is the
$\mu$-polarized electron number fluctuation term.

Using again the quasi-equilibrium condition, Eq. (\ref{quasi}), we
obtain \br\label{nc1} \frac{d}{dt}n_{e{\bf
k}}^\mu&=&\Lambda_{e{\bf k}}^\mu(1-n_{e{\bf
k}}^\mu)-\gamma_{nr}^\mu n_{e{\bf k}}^\mu
-\sum_{ll'\mu'}[{\cal D}_{l{\bf k}\mu\mu'}g_{l{\bf k}\mu\mu'}g^*_{l'{\bf k}\mu\mu'} A_{l\mu\mu'}^\dagger A_{l'\mu\mu'}(n_{e{\bf k}}^\mu+n_{h-{\bf k}}^{\mu'}-1)+H.c.]\nonumber\\
&&+\sum_{l\mu'}(i{\cal D}_{l{\bf k}\mu\mu'}g^*_{l{\bf k}\mu\mu'}
A_{l\mu\mu'}^\dagger F_{\sigma_{\bf k}}^{\mu\mu'}+H.c.)+ F_{e{\bf
k}}^\mu. \er%
This last equation can be further simplified by neglecting
correlation between modes, such that \br\label{nc}
\frac{d}{dt}n_{e{\bf k}}^\mu&=&\Lambda_{e{\bf k}}^\mu(1-n_{e{\bf
k}}^\mu)-\gamma_{nr}^\mu n_{e{\bf k}}^\mu
-\sum_{l\mu'}(G_{{\bf k} ll }^{\mu\mu'}+G_{{\bf k} ll }^{*\mu\mu'}) n_{l\mu\mu'}\nonumber\\
&&+\sum_{l\mu'}(i{\cal D}_{l{\bf k}\mu\mu'}g^*_{l{\bf k}\mu\mu'}
A_{l\mu\mu'}^\dagger F_{\sigma_{\bf k}}^{\mu\mu'}+H.c.)+ F_{e{\bf
k}}^\mu. \er%
 Since the third term of the right hand side of Eq. (\ref{nc}) is
due to the radiative recombination we can simplify it by just
relating it to the radiative decay rate as follows
  \cite{sargent,haug}
\br\label{nc2} \frac{d}{dt}n_{e{\bf k}}^\mu&=&\Lambda_{e{\bf
k}}^\mu(1-n_{e{\bf k}}^\mu)-\gamma^\mu_{nr} n_{e{\bf k}}^\mu
-\gamma_{r}^\mu n_{e{\bf k}}^\mu+ F_{e{\bf k}}^\mu, \er where we
have also included the fourth term of Eq.(\ref{nc}) in the
definition of $F_{e{\bf k}}^\mu$, and obviously, $\gamma_r^\mu$ is
a carrier occupation number dependent function as \br
\gamma_r^\mu=\frac{\sum_{l\mu'}(G_{{\bf k} ll }^{\mu\mu'}+G_{{\bf
k} ll }^{*\mu\mu'}) n_{l\mu\mu'}}{n_{e{\bf k}}^\mu},\er and
$G_{{\bf k} ll }^{\mu\mu'}$ is also an implicit function of
$n_{e{\bf k}}^{\mu}$. Depending on the process involved in the
non-radiative recombination, $\gamma_{nr}^\mu$ can also be
$n_{e{\bf k}}^{\mu}$-dependent. For simplicity we have taken both,
the radiative and non-radiative recombination rates as constants,
and as such independent of the magnetic field. In this regime the
average value for the carrier number is given as a function of the
pumping rate as
\begin{equation}
\langle n_{e{\bf k}}^\mu(t)\rangle=\left(\langle n_{e{\bf
k}}^\mu(0)\rangle-\frac{\Lambda_{e{\bf k}}^\mu} {\Lambda_{e{\bf
k}}^\mu+\gamma_{nr}^\mu+\gamma_r^\mu}\right) e^{-(\Lambda_{e{\bf
k}}^\mu+\gamma_{nr}^\mu+\gamma_r^\mu)t}+\frac{\Lambda_{e{\bf
k}}^\mu} {\Lambda_{e{\bf k}}^\mu+\gamma_{nr}^\mu+\gamma_r^\mu},
\end{equation}
whose stationary solution is
\begin{equation}\label{enum}
\langle n_{e{\bf k}}^\mu\rangle_{eq.}=\frac{\Lambda_{e{\bf k}}^\mu}
{\Lambda_{e{\bf k}}^\mu+\gamma_{nr}^\mu+\gamma_r^\mu}.
\end{equation}
Similarly the equilibrium hole occupation probability is given by,
\begin{equation}\label{hnum}
\langle n_{h-{\bf k}}^{\mu'}\rangle_{eq.}=\frac{\Lambda_{h-{\bf k}}^{\mu'}}
{\Lambda_{h-{\bf k}}^{\mu'}+\gamma_{nr}^{\mu'}+\gamma_r^{\mu'}},
\end{equation}
where $\Lambda_{h-{\bf k}}^{\mu'}$ is the hole pumping rate and
$\gamma_{nr}^{\mu'}$ and $\gamma_r^{\mu'}$ are the non-radiative
and radiative hole recombination rate, respectively. Thus the
expected spontaneous emission rate (\ref{spont}) can be simply
given by
\begin{equation}\label{emmrate}
\langle R_{sp,l}^{\mu\mu'}\rangle=\frac\gamma 2 \sum_{\bf
k}|g_{l{\bf k}\mu\mu'}|^2{\cal L}_{l{\bf k}}^{\mu\mu'}
\frac{\Lambda_{e{\bf k}}^\mu\Lambda_{h-{\bf k}}^{\mu'}}
{(\Lambda_{e{\bf k}}^\mu+\gamma_{nr}^\mu+\gamma_r^\mu)
(\Lambda_{h-{\bf k}}^{\mu'}+\gamma_{nr}^{\mu'}+\gamma_r^{\mu'})}.
\end{equation}
To use this last expression, it is convenient to write the
spectral light polarization
 as given by Eq. (\ref{poldeg}) in the following compact form
\begin{equation}\label{poldeg2}
{\cal P}(\nu_l)=\frac{\sum_{\mu\mu'}(\mu-\mu')\langle
R_{sp,l}^{\mu\mu'}\rangle} {\left[\sum_{\mu\mu'}\langle
R_{sp,l}^{\mu\mu'}\rangle+4\bar n_0 (\nu_l)\right]},
\end{equation} where we must remember that
the elements $R_{sp,l}^{-\frac 1 2 -\frac 1 2}= R_{sp,l}^{-\frac 1
2 \frac 3 2}=R_{sp,l}^{\frac 1 2 -\frac 3 2}= R_{sp,l}^{\frac 1 2
\frac 1 2}=0$. Substituting Eq. (\ref{emmrate}) into Eq.
(\ref{poldeg2}) the spectral light polarization is finally given
in function of the balance of electron and hole injection as
\begin{equation}\label{poldeg3}
{\cal P}(\nu_l)=\frac{\sum_{{\bf k}\mu\mu'}(\mu-\mu')|g_{l{\bf
k}\mu\mu'}|^2{\cal L}_{l{\bf k}}^{\mu\mu'} \Gamma_{\bf
k}^{\mu\mu'} } {\left[\sum_{{\bf k}\mu\mu'}|g_{l{\bf
k}\mu\mu'}|^2{\cal L}_{l{\bf k}}^{\mu\mu'} \Gamma_{\bf
k}^{\mu\mu'} +8\bar n_0 (\nu_l)/\gamma\right]}, \end{equation}
where we have defined
\begin{equation}\label{gamma}
\Gamma_{\bf k}^{\mu\mu'}\equiv \frac{\Lambda_{e{\bf
k}}^\mu\Lambda_{h-{\bf k}}^{\mu'}} {(\Lambda_{e{\bf
k}}^\mu+\gamma_{nr}^\mu+\gamma_r^\mu) (\Lambda_{h-{\bf
k}}^{\mu'}+\gamma_{nr}^{\mu'}+\gamma_r^{\mu'})}, \end{equation} as
the pumping to recombination rate. As before the light degree of
polarization is given by integrating (\ref{poldeg3}).

\subsection{Pumping rate modelling}
Before proceed further we need to discuss the phenomenologically
introduced pumping-rate in detail. When summing over ${\bf k}$ the
pumping and pump blocking term for the $\alpha$-carrier
($\alpha=e$, or $h$), must be related to the spin polarized
current density $J_\mu$ \cite{sargent,haug} by
\begin{equation} \sum_{{\bf k}}\Lambda_{\alpha{\bf
k}}^\mu(1-n_{e{\bf k}}^\mu)=\frac{\eta J_\mu}{e d},\end{equation}
where $\eta$ is the total quantum efficiency that the injected
carriers contribute to the population of the $\alpha\mu$-subband,
$e$ is the electron charge and $d$ is the thickness of the active
region. Assuming that by the time the injected carriers reach the
active region they collide often enough to be in equilibrium
within each subband, it is reasonable to assume the
quasi-equilibrium condition \cite{sargent,haug} such that
\begin{equation}\label{lambda}\Lambda_{\alpha{\bf k}}^\mu=\frac{\eta_{tr} J_\mu}{e d N_0}
f_{\alpha{\bf k}0},\end{equation}
 where $N_0$ and
$f_{\alpha{\bf k}0}$ are the total carriers density and the
Fermi-Dirac distribution function, respectively, at zero bias.
$\eta_{tr}$ is the transport part of the quantum efficiency,
giving the efficiency that the injected carriers reach the active
region. $\eta_{tr}$ could include a spinorial dependence to take
into account dephasing and decoherence mechanisms at the
spin-aligner material and GaAs interface \cite{wu}. However such
mechanisms are not concerned in the present work. The spin
dependent current density $J_\mu$ is related to the spin-alignment
efficiency of the material cap layer (Fig. 1). Spin-aligner
materials such as Be$_{1-x-y}$Mn$_x$Zn$_{y}$Se \cite{fiederling},
Zn$_{1-x}$Mn$_{x}$Se \cite{jonker} or ferromagnetic GaMnAs
epilayers \cite{ohno} show giant magnetoresistance
\cite{furdyna,fukumura}. Thus $J_\mu$ must take into account the
magnetic field strength relating spin aligned carrier injection
into the GaAs LED. From Refs.
\cite{fiederling,ohno,jonker,jonker2,ploog} the spin aligned
current injection follows closely the profile of a Brillouin
paramagnet, whose net magnetization is phenomenologically given by
\cite{furdyna}
\begin{equation}
M=\frac{\overline{x}}{x}{\cal G}_\alpha^\prime \mu_B S \;{\cal
B}_S\left(\frac{{\cal G}_\alpha^\prime\mu_B S B}{k_B
(T+T_0)}\right)
\end{equation}
where ${\cal G}_\alpha^\prime$ is the magnetic material electronic
g-factor, S is the magnetic material spin, ${\cal B}_S$ is a
$S$-Brillouin function and $\frac{\overline{x}}{x}$ is the molar
fraction of Mn contributing to the saturation of the magnetization
and $T_0$ is a fitting temperature  to scale with the experimental
magnetization curve \cite{furdyna}. Since the degree of
polarization of the injected current is directly proportional to
the magnetization and also directly proportional to the magnetic
semiconductor layer thickness $d_{ms}$, we assume the following
phenomenological electronic injection current density
\begin{equation}\label{curr}
J_\mu=\frac{J_0}{2}+ d_{ms}\frac{\overline{x}}{x}{\cal G}_e^\prime
\mu_B \mu \;{\cal B}_{1/2}\left(\frac{{\cal G}_e^\prime\mu_B  B
d_{ms}}{2 k_B Td_0}\right),
\end{equation}
where $J_0$ is the net current density without a magnetic field.
The net current is always $J_0$, but each component of $J_\mu$ is
increased or decreased if $\mu=1/2$ or $-1/2$, respectively.
Remark that instead of $T_0$ we included the fraction $d_{ms}/d_0$
as a fitting parameter, where $d_0$ is a fitting length, which is
more convenient for our purposes. If we define the polarization of
the injected current by
\begin{eqnarray}
P_j&\equiv&\frac{J_{\frac 1 2}-J_{-\frac 12}}{J_{\frac 1
2}+J_{-\frac 12}},\end{eqnarray} which is the rate between spin
and charge current densities, we obtain by Eq. (\ref{curr})
\begin{eqnarray}P_j=\frac{1}{J_0}d_{ms}\frac{\overline{x}}{x}{\cal G}_e^\prime \mu_B
\;{\cal B}_{1/2}\left(\frac{{\cal G}_e^\prime\mu_B  B d_{ms}}{2k_B
Td_0}\right),
\end{eqnarray}
which then shows a Brillouin function dependence with the magnetic
field, the inverse of the temperature, as well as a linear
dependence with the spin-aligner material thickness as observed
experimentally \cite{fiederling,jonker,jonker2,ploog}. Notice that
instead of including the temperature dependence in the magnetic
semiconductor g-factor we have assumed this dependence in the
phenomenological magnetization \cite{furdyna}. In Fig. 6 we plot
the normalized polarization $P_j^*=P_jJ_0 x/\overline{x}d_{0}{\cal
G}_e^\prime \mu_B$, i.e., $(d_{sm}/d_0){\cal
B}_{1/2}\left(\frac{{\cal G}_e^\prime\mu_B B}{2k_B T}\right)$, as
function of the magnetic field in Fig. 6a and the temperature in
Fig. 6b. These figures clearly show the observed injection
polarization \cite{fiederling} by varying $B$, $T$ and the
magnetic semiconductor spin-aligner thickness, justifying our
pumping rate modelling through Eqs.
(\ref{lambda}) and (\ref{curr}).

\vspace{-0.4cm}
\begin{figure}
\centerline{$\;$\hskip 0
truecm\psfig{figure=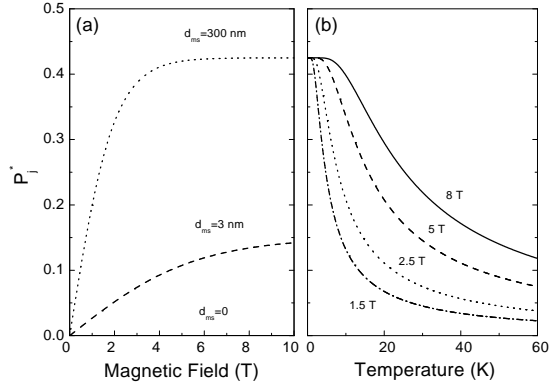,height=12cm,angle=0}}\vspace{-5cm}
\centerline{ \caption{Normalized polarization of injected carriers
into LED from a Brillouin magnetic semiconductor. (a) Carrier
injection polarization dependence with the applied magnetic field
and magnetic semiconductor thickness $d_{ms}$ at T=4.2 K. (b)
Carrier injection polarization dependence with the temperature for
$d_{ms}=300$ nm.} } \label{fig6}
\end{figure}

\subsection{Net light emission polarization}
Now we can include the spin-aligned carrier injection, as
described above, in the polarized light emission (\ref{poldeg3}).
The spin aligned carrier injection reflects as an unbalanced
carrier population through Eq. (\ref{lambda}). We must remark that
due to the reduced spin-orbit coupling in the conduction band,
spin-injection of electrons is more efficient than holes. Thus we
simply set a balanced constant pumping rate for holes from the
drain lead, while considering an electronic spin-aligned
injection. For the following calculations we fixed the temperature
to T = 4.2 K, where the thermal photons emission is negligible,
and thus can be simply disregarded from Eq. (\ref{poldeg3}).

In working device regime radiative recombination rate is always
much higher than nonradiative recombination rate as the former is
dominant and the latter is a disturbance due to the impurities and
other undesirable material defects. Thus non-radiative
recombination rate is always smaller than pumping rate, even in
the low injection limit. The radiative recombination rate,
however, play a crucial role for the limiting regimes for the pump
to recombination rate (\ref{gamma}). First let us consider the
regime of strong pumping rate where
$\gamma_r^\mu\ll\Lambda_{^\alpha{\bf k}}^\mu$. Thus $\Gamma_{\bf
k}^{\mu\mu'}$ saturates to $1$ and (\ref{poldeg3}) simplifies to
\begin{equation}\label{poldeg4}
{\cal P}(\nu_l)=\frac{\sum_{{\bf k}\mu\mu'}(\mu-\mu')|g_{l{\bf
k}\mu\mu'}|^2{\cal L}_{l{\bf k}}^{\mu\mu'} } {\sum_{{\bf
k}\mu\mu'}|g_{l{\bf k}\mu\mu'}|^2{\cal L}_{l{\bf k}}^{\mu\mu'}},
\end{equation}
which is a saturation for the emitted light polarization, since in
that limit all the electronic and hole states are occupied, as
follows from Eqs. (\ref{enum}) and (\ref{hnum}), leaving no free
state for carrier injection. The polarization is then dependent
only on the spectral shape of the GaAs light emission and
corresponds to the intrinsic emission we studied before, in the
limit of high occupancy. On the other hand, for the regime of weak
pumping, when $\gamma_r^\mu\gg\Lambda_{^\alpha{\bf k}}^\mu$, the
pumping to recombination rate reads
\begin{equation}\label{gamma}
\Gamma_{\bf k}^{\mu\mu'}\equiv \frac{\Lambda_{e{\bf
k}}^\mu\Lambda_{h-{\bf k}}^{\mu'}} {(\gamma_{nr}^\mu+\gamma_r^\mu)
(\gamma_{nr}^{\mu'}+\gamma_r^{\mu'})}\ll 1,
\end{equation}which is the limit
where all the electronic and hole states are almost unoccupied,
due to the fast recombination process. In this situation the net
light emission polarization is then strongly dependent on the
polarized carrier injection, but with the GaAs light emission
features. This limit is also consistent with the low injection
limit we have taken before. The net spectral polarization is then
given by
\begin{equation}\label{poldeg5}
{\cal P}(\nu_l)=\frac{\sum_{{\bf k}\mu\mu'}(\mu-\mu')|g_{l{\bf
k}\mu\mu'}|^2{\cal L}_{l{\bf k}}^{\mu\mu'}f_{e{\bf
k}0}\left[J_0/2+\mu d_{ms}(\bar x/x){\cal G}_e^\prime \mu_B\;{\cal
B}_{1/2}\left(\frac{{\cal G}_e^\prime\mu_B  B d_{ms}}{2k_B
Td_0}\right)\right] } {\sum_{{\bf k}\mu\mu'}|g_{l{\bf
k}\mu\mu'}|^2{\cal L}_{l{\bf k}}^{\mu\mu'}f_{e{\bf
k}0}\left[J_0/2+\mu d_{ms}(\bar x/x){\cal G}_e^\prime \mu_B\;{\cal
B}_{1/2}\left(\frac{{\cal G}_e^\prime\mu_B  B d_{ms}}{2k_B
Td_0}\right)\right]},
\end{equation}
from where we obtain by integration the net light emission
polarization as plotted in Fig. 7 by varying $B$ and $d_{ms}$. Due
to the low value of the Land\'e g-factor for electrons in GaAs,
the Zeeman splitting
 is very small, but it is contrary to the
splitting of the spin-aligner material, decreasing the
polarization, which contributes to the decreasing of the
saturation value for the net degree of polarization. Both the
polarization of the spin-injected electrons ($P_j$) and that due
to intrinsic g-factor ($P$) increase in magnitude as the applied
magnetic field increases, being however $P_j$ opposite to $P$. In
our model $P_j$ is dominant up to 7 T, where $P_j$ saturates but
$P$ does not, therefore the net polarization drops as evidenced
experimentally \cite{fiederling,jonker,jonker2} (see the inset of
Fig. 7). The spin-aligner material layer thickness is also an
important feature for the net degree of polarization. Remark that
for this figure we have considered both light and heavy hole
states, and thus the highest polarization possible to be attained
is 50$\%$. Had we neglected light-hole states the polarization
could be as high as 100$\%$ depending on the spin-aligner material
layer thickness. For $d_{ms}=300$ nm the higher attained
efficiency of polarization would be approximately 85$\%$, which is
in complete agreement with the observed value of 86$\%$ from
Ref.\cite{fiederling}.

\vspace{-1cm}
\begin{figure}
\centerline{$\;$\hskip 0
truecm\psfig{figure=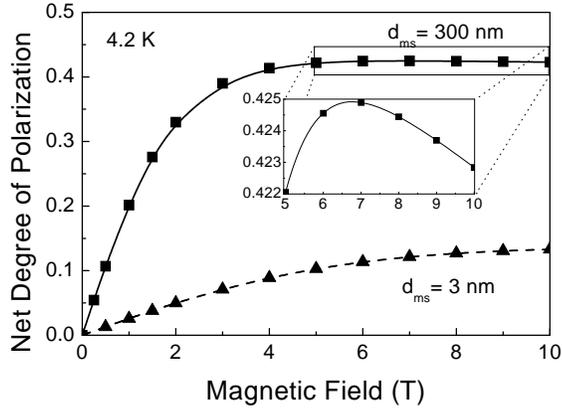,height=12cm,angle=0}} \vspace{-5cm}
\centerline{ \caption{Net light polarization with inclusion of
spin-aligned carrier injection. The degree of light polarization
is dependent on the magnetic material layer thickness. The inset
show the decrease of the saturated spin-aligned due to the
intrinsic GaAs polarized light emission.} } \label{fig7}
\end{figure}
\section{Concluding remarks}
In conclusion we have shown that the Langevin approach is quite
useful for the microscopic description of spin-mediated polarized
light emission. We have quantified the intrinsic degree of
polarization of the GaAs light emission, being it strongly
affected by temperature effects. We have shown that the
temperature dependence of the electronic g-factor is responsible
for a slight decrease of the degree of polarization, once the
decrease of the electronic g-factor with the temperature decreases
the conduction band spin-splitting sensitivity to the magnetic
field. However at higher temperatures, thermal photons are also
emitted by the GaAs device, and the intrinsic degree of
polarization decreases abruptly at the threshold temperature
($T_c$). The effect of unbalanced spin-injection was also analyzed
reflecting the dependence on the spin-aligned carrier pumping, as
well as on the radiative and the non-radiative electron-hole
recombination. Since the intrinsic polarization in GaAs is
opposite to that in spin-polarizing materials, it decreases the
net spin-injection efficiency as reported in
\cite{fiederling,jonker,jonker2}. We have modelled the
spin-polarized carrier injection by considering the spin-aligner
as a Brillouin paramagnet\cite{furdyna}, and introduced a
phenomenological spin-polarized current density, which is
dependent on the spin-aligner layer thickness,
 the applied magnetic field and the temperature as well.

 As a final remark, throughout this paper we have assumed the dipole quasi-equilibrium
 regime for analyzing the light emission polarization. That means
 we have considered that each electronic spin component is in equilibrium
  inside each sub-band when radiative processes take
 place. This is actually the situation for working devices regime.
 However the non-equilibrium regime, where the dipole dephasing and decoherence rate are
  taken into
  account, is interesting for the treatment of optical detection
   of spin relaxation processes \cite{oestreich3}. The formalism here developed
   can be readily applied to these problems and could bring some
   enlightening on the microscopic mechanism related to spin relaxation
   in semiconductors media.

\acknowledgments{MCO thanks G.-Q. Hai and G. A. Prataviera for
enlightening discussions. MCO was supported by FAPESP,
 under projects 01/00530-2 and 00/15084-5}
%

\newpage

{\bf FIGURE CAPTIONS}

\bigskip

{\bf Fig 1.}  Spin-filtering device.

\bigskip

{\bf Fig 2.}  Radiative inter-band transitions allowed in GaAs.

\bigskip

{\bf Fig 3.}  Intrinsic electroluminescence spectra of GaAs as
function of the magnetic field.

\bigskip

{\bf Fig 4.}  Intrinsic degree of polarization of GaAs in function
of the magnetic field.

\bigskip

{\bf Fig 5.}  Decreasing of intrinsic polarization of GaAs as a
function of the temperature.

\bigskip

{\bf Fig 6.}  Normalized polarization of injected carriers into
LED from a Brillouin magnetic semiconductor. (a) Carrier injection
polarization dependence with the applied magnetic field and
magnetic semiconductor thickness $d_{ms}$ at T=4.2 K. (b) Carrier
injection polarization dependence with the temperature for
$d_{ms}=300$ nm.

\bigskip

{\bf Fig 7.} Net light polarization with inclusion of spin-aligned
carrier injection. The degree of light polarization is dependent
on the magnetic material layer thickness. The inset show the
decrease of the saturated spin-aligned due to the intrinsic GaAs
polarized light emission.

\begin{references}
\bibitem{footnote} Remark that although the employed model be considered
 inside the working device
regime, i.e. for ``good" devices, where the radiative
recombination rate is always much higher than nonradiative
recombination rate due to phonon scattering, the observed behavior
is still valid outside this regime. Phonon scattering will cope
with the radiative thermal effect on decreasing the intrinsic
light emission degree of polarization.
\end{references}
\end{document}